\documentstyle[epsfig]{aipproc}

\def\fun#1#2{\lower3.6pt\vbox{\baselineskip0pt\lineskip.9pt
        \ialign{$\mathsurround=0pt#1\hfill##\hfil$\crcr#2\crcr\sim\crcr}}}

\renewcommand\({\left(}
\renewcommand\){\right)}
\renewcommand\[{\left[}

\newcommand\eq[1]{Eq.~(\ref{#1})}

\newcommand\ee{\end{equation}}
\newcommand\be{\begin{equation}}
\newcommand\eea{\end{eqnarray}}
\newcommand\bea{\begin{eqnarray}}

\newcommand\GeV{\,\mbox{GeV}}

\newcommand\mpl{M_{\rm P}}

\newcommand\lsim{\mathrel{\rlap{\lower4pt\hbox{\hskip1pt$\sim$}}
    \raise1pt\hbox{$<$}}}
\newcommand\gsim{\mathrel{\rlap{\lower4pt\hbox{\hskip1pt$\sim$}}
    \raise1pt\hbox{$>$}}}

\def\dslash{\not{\hbox{\kern-2pt $\partial$}}}
\def\Dslash{\not{\hbox{\kern-4pt $D$}}}
\def\Oslash{\not{\hbox{\kern-4pt $O$}}}
\def\Qslash{\not{\hbox{\kern-4pt $Q$}}}
\def\pslash{\not{\hbox{\kern-2.3pt $p$}}}
\def\kslash{\not{\hbox{\kern-2.3pt $k$}}}
\def\qslash{\not{\hbox{\kern-2.3pt $q$}}}

 \newtoks\slashfraction
 \slashfraction={.13}
 \def\slash#1{\setbox0\hbox{$ #1 $}
 \setbox0\hbox to \the\slashfraction\wd0{\hss \box0}/\box0 }
 
\def\ee{\end{equation}}
\def\be{\begin{equation}}

\def\calp{{\cal P}}

\newcommand\sub[1]{_{\rm #1}}

\begin{document}

\title{Models of inflation and their predictions
}

\author{David H. Lyth}
\address{Department of Physics,\\
Lancaster University,\\
Lancaster LA1 4YB.~~~U.~K.}

\maketitle

\begin{abstract}
Taking field theory seriously, inflation model-building is difficult
but not impossible. The observed value of the spectral index 
of the adiabatic density perturbation is starting to discriminate 
between models, and may well pick out a unique one in the forseeable 
future.
\end{abstract}

I shall summarise the present status of inflation model-building,
and its comparison with observation.
This is already a substantial area of 
research, and 
with the advent of new observations in the next few years
it will become a major industry. For an extensive review with
 references, see \cite{review99}.

I focus on the simplest paradigm, which following the usual scientific 
practice should be tested to destruction before complications are 
entertained. There is a 
{\em slowly-rolling, single-component} inflaton field, which experiences
{\em Einstein gravity} and drives the observable Universe
into a {\em spatially flat} condition. 
The {\em gaussian adiabatic} density perturbation,
generated by the vacuum fluctuation of the inflaton field $\phi$, is solely
responsible for the origin of structure.

During inflation, the potential $V(\phi)$ of the inflaton field $\phi$
satisfies the 
flatness conditions 
\be
\epsilon \ll 1,\hspace{5em} |\eta| \ll 1 \,, \label{flat}
\ee
where $\epsilon \equiv \frac12\mpl^2(V'/V)^2$ and
$\eta \equiv \mpl^2 V''/V$.
The inflaton field satisfies the slow-roll approximation
$3H\dot\phi=-V'$ where $H$ is the Hubble parameter 
given by $3H^2=V/\mpl^2$. To work out the predictions one needs 
the number of 
$e$-folds $N$ between a given epoch and the end
of slow-roll inflation (with the inflaton $\phi$). 
Its small change is defined by $dN  \equiv -H\,dt(=-d\ln a)$,
which with the slow-roll approximation leads to
\be
N(\phi) = \int^\phi_{\phi_{\rm end}} \mpl^{-2}\frac V{V'} d\phi \,.
\label{nint}
\label{Nrelation}
\label{dndphi}
\ee
Here $\phi\sub{end}$ marks the end of slow-roll inflation, 
caused by the failure of the flatness conditions or by
the destabilization of a non-inflaton field. Often,
the integral is dominated by the other limit $\phi$ in which case
the predictions are independent of
$\phi\sub{end}$.

The vacuum fluctuation of the inflaton field
generates a gaussian adiabatic primordial density perturbation, whose 
conventionally-defined spectrum is given by
\be
\delta_H^2(k) = \frac1{75\pi^2\mpl^6}\frac{V^3}{V'^2} \,.
\label{delh}
\ee
The right hand side is evaluated at the value $\phi(k)$ which
corresponds to horizon exit
$k=aH$. It satisfies
$d\ln k = -dN(\phi)$
and therefore $\ln(k\sub{end}/k)= N(\phi)$,
where $k\sub{end}$ is the scale leaving the horizon at the end of
slow-roll inflation. With \eq{nint}, this determines $\phi(k)$
provided that we know the value of $N(\phi)$ when some reference
scale leaves the horizon. This scale is conveniently taken to be the
central scale probed by COBE,
$k\sub{COBE}\simeq 7.5 H_0$ where $H_0$ is the present Hubble
parameter. Depending on the history of the Universe, one has
\be
N\sub{COBE} \simeq 60 - \ln(10^{16}\GeV/V^{1/4}) - \frac13\ln(V^{1/4}/T\sub{
reh})
-\Delta N
\,,
\ee
where $T\sub{reh}$ is the reheat temperature and $\Delta N>0$ allows
for matter domination and thermal inflation between reheating and 
nucleosynthesis (and any continuation of inflation after the epoch
$\phi\sub{end}$).

Differentiating \eq{delh} with the aid of \eq{nint}, the
spectral index is 
\be
\frac{n(k)-1}2 \equiv \frac{d\delta_H}{d\ln k}
= \eta-3\epsilon \,.
\label{n2}
\ee
{\em If} $n$ is constant then $\delta_H^2 \propto k^{n-1}$.

Inflation also generates gravitational waves with 
primordial spectrum $
\calp\sub{grav} (k) = \frac 2{\mpl^2} \left( \frac H{2\pi} \right)^2$.
No gravitational wave signal is seen in the cmb anisotropy,
which translates into a bound
$\epsilon \lsim 0.1$. The signal will probably never be seen unless
$\epsilon\gsim 10^{-3}$.

At $k\sub{COBE}$, 
the COBE observations give the accurate normalization
(ignoring gravitational waves)
$\delta_H=1.91\times 10^{-5}$, 
which corresponds to
\be
V^{1/4}/\epsilon^{1/4}=.027\mpl=6.7\times 10^{16}\GeV.
\label{vbound}
\label{cobenorm}
\ee
The present bound $\epsilon\lsim 0.1$ on gravitational waves implies 
$V^{1/4}\lsim 3.6\times 10^{16}\GeV$. Gravitation waves will never be 
detectable if $V^{1/4}\lsim 1\times 10^{16}\GeV$, and 
{\em most inflation 
models give a lower value} when normalized to satisfy \eq{vbound}.

Over the range  of cosmological scales, say
$H_0 < k < 10^4 H_0$, there 
is an observational bound on the scale-dependence of
$\delta_H$. Until recently uncertainties in the cosmological parameters
allowed only the weak result 
$|n-1|\lsim 0.2$, but new data give a preliminary result $|n-1|<0.05$
\cite{dick}. After Planck flies we shall probably know
$n(k)$ with an uncertainty of $\pm 0.01$.

At the most primitive level, a model of inflation consists 
of a form for
$V(\phi)$, plus a prescription for $\phi\sub{end}$ if the latter is not 
determined by $V$ as happens in some hybrid inflation models.
 From a field theory viewpoint, one expects $V(\phi)$ to be 
schematically of the following form\footnote{The form is 
more restrictive if $\phi$ is a pseudo-Goldstone boson, but that 
hypothesis has not so far lead to an attractive model of inflation.}
\bea
V&=&V_0 + \frac12m^2\phi^2 + M\phi^3+\frac14\lambda\phi^4 \nonumber\\
&+& (\tilde m^4 +2g\phi^2+g^2\phi^4)\ln(g\phi/Q) \nonumber\\
&+& \sum_{d=5}^\infty \lambda_d \mpl^{4-d} \phi^d \nonumber \\
 &+& [\Lambda^{4+\alpha} \phi^{-\alpha} - \tilde\Lambda^{4\pm\beta}
\phi^{\mp \beta} ]
 \label{vexp} 
\,.
\eea
In the first line are {\em renormalizable 
tree-level} terms, with the origin is chosen
so that $V'=0$; the coefficients can have either sign. In the second line is 
the {\em one-loop correction} due to a particle with mass $\tilde m$
and coupling $g$, valid if $g\phi\gsim \tilde m$. 
(It is suppressed at smaller $\phi$.) The renormalization scale $Q$ 
should be fixed at a typical relevant value of $g\phi$
to minimize higher loop contributions.
One sums over particles with a plus/minus sign for 
bosons/fermions, and unbroken 
global supersymmetry would make the total vanish. During inflation
susy is broken, but the $\phi^4$ term still vanishes, and the
$\phi^2$ term {\em may} vanish, but one expects no cancellations between 
the contributions to the the constant term.
The third line contains the
{\em non-renormalizable} terms which 
summarise unknown Planck scale physics; the coefficients $\lambda_d$
are generically of order
1, but supersymmetry can make a finite number of them tiny.
(By appealing to a continuous global symmetry it can make them
all tiny, but no such symmetry comes out of string theory.)
The fourth line contains a $\phi^{-\alpha}$ term that might come from
dynamical symmetry breaking, and a $-\phi^{\mp\beta}$ term that might
come from mutated hybrid inflation. These terms will be present only
in exceptional cases, unlike the others which are generic.

One who presumes to use
field theory ought to take this expression seriously, and when that is 
done the flatness conditions \eq{flat} turn out to be extraordinarily 
difficult to satisfy. The non-renormalizable terms are obviously 
dangerous. So are the loop corrections, especially
in the context of hybrid inflation where some coupling has to be 
substantial. Less obviously, a generic supergravity theory
gives a prediction of the form $\mpl^2 V''/V = 1+\cdots$, in which case
there has to be some cancellation whose origin is at present obscure.

The simplest proposed model, usually called chaotic inflation,
is a monomial
$V\propto \phi^p$ with $p$ usually 2 or 4.
Inflation takes place at $\phi>\phi\sub{end}
\sim p\mpl$, giving $n-1 = -(2+p)/(2N)$ and significant 
gravitational waves ($\epsilon=p/(5N)$).

If non-renormalizable terms are there, they kill the above model. To 
live with them one needs $\phi\lsim\mpl$ or 
$\phi \ll \mpl$. (The latter case is preferable, but one has to 
watch the loop correction which generates a term $V''\propto
\phi^{-2}$.) Making the reasonable assumption that
only one term of \eq{vexp} is relevant, \eq{flat} then requires 
$V\simeq V_0$.

Predictions for the spectral index are given in the
Tables. An inflaton field 
with negligible interaction ($V=V_0\pm\frac12m^2\phi^2
$) gives a constant $n-1$, which can be positive or negative and is 
typically not extremely 
small or interactions {\em would} be significant.
One significant interaction term typically gives 
$n$ close to 1, with weak scale-dependence.

A dramatic exception is the 
case \cite{ewanloop} where a $\phi^2\ln \phi$ loop correction dominates
the mass term, as shown in the second line of Table 1.
The correct COBE normalization \eq{vbound} is obtained with a
reasonable value $c\sim 10^{-1}$ to $10^{-2}$ of the coupling $c$.\footnote
{To calculate this normalization one has to take into account higher 
loop corrections by using a renormalization-group-improved 
potential, but the order of magnitude is unaffected.}
Furthermore, such a coupling allows $n$ to pass through 1 on cosmologically 
interesting scales!

The observed value of $n(k)$ will become an increasingly powerful
discriminator in the future. If one were to take it 
seriously, the preliminary result 
$|n-1|<0.05$ would already rule out
the cubic self-interaction in Table 2.
It would also strongly constrain the parameters $c$ and $\sigma$
in the case just mentioned, perhaps demanding physically unreasonable
values for them.

\begin{table}
\centering
\caption{Predictions for the spectral index $n(k)$, using $N=
\ln(k\sub{end}/k)$ where $k\sub{end}=aH$ at the end of slow-roll
inflation. Constants $c$, $q$ and $A$ are positive 
while $\sigma$ and $p$ can have either sign.
In the first two cases one expects $|c|\sim 10^{-1}$ to $10^{-2}$,
and in the second case, one expects $|\sigma|\gsim |c|$.}
\begin{tabular}{|llll|}
\noalign{\vspace{-8pt}} 
Comments 
& $V(\phi)/V_0\simeq 1$ & $\frac12 (n-1)$ & $\frac12\frac{dn}{d\ln k}$
\\[4pt] \hline
Mass term & $1\pm\frac12c \frac{\phi^2}{\mpl^2} $ 
& $\pm c$ & 0 
\\[4pt]
Softly broken susy &
$1\pm\frac12c \frac{\phi^2}{\mpl^2} \ln \frac\phi A$ &
$\pm c + \sigma e^{\pm c N} $& $\mp c \sigma e^{\pm c N}$ \\[4pt]
Spont. broken susy &
$1+c \ln\frac\phi A$ & $-\frac1{2N} $ &  $-\frac12\frac1{N^2}$ \\[4pt]
$p>2$ or $-\infty<p<1$ & $1-c\phi^p$ &
$-\(\frac{p-1}{p-2} \) \frac1 N$   &  
$-\(\frac{p-2}{p-1}\) \( \frac{n-1}2 \)^2$ \\[4pt]
$p$ integer $\leq -1$ or $\geq 3$ & $1+c\phi^p$ &
$\frac{p-1}{p-2}\frac1{N\sub{max}-N}$ & 
$-\(\frac{p-2}{p-1}\) \( \frac{n-1}2 \)^2$ \\[4pt]
\hline
\end{tabular}
\end{table}

\begin{table}
\caption[table2]{\label{t:2} Some predicted values.}
\begin{tabular}{|cllll|}
\noalign{\vspace{-8pt}} $V(\phi)/V_0$ 
& \multicolumn{2}{c}{$1-n$} & \multicolumn{2}{c|}{$-10^3dn/d\ln k$} \\
& $N=50$ & $N=20$ & $N=50$ & $N=20$  \\ \hline
$1+c\ln(\phi/Q)$ & $0.02$ & $0.05$ & $(0.4)$ & $2.6$ \\
$1-c\phi^{-2}$ & $0.03$ & $0.075$ & ($0.6$) & $3.8$ \\
$1-c\phi^4$ & $0.06$ & $ 0.15$ & ($1.2$) & $5.4$ \\
$1-c\phi^3$ & $0.08$ & $ 0.20$ & ($1.6$) & $10.0$ \\ \hline
\end{tabular}
\end{table}

\end{document}